\documentclass[a4paper,11pt]{article}
\pdfoutput=1 

\usepackage{jheppub} 

\usepackage[T1]{fontenc} 
\usepackage{physics, tensor}
\usepackage{amssymb, amsmath}

\title{Are there Einsteinian gravities involving covariant derivatives of the Riemann tensor?}  

\author[a,b]{Jos\'e D. Edelstein,}
\author[a,b]{Alberto Rivadulla S\'anchez}
\author[a,b]{and David V\'azquez Rodr\'\i guez}
\affiliation[a]{Departamento de F\'\i sica de Part\'\i culas, Universidade de Santiago de Compostela, E-15782 Santiago de Compostela, Spain}
\affiliation[b]{Instituto Galego de F\'\i sica de Altas Enerx\'\i as (IGFAE), Universidade de Santiago de Compostela, E-15782 Santiago de Compostela, Spain}
\emailAdd{jose.edelstein@usc.es}
\emailAdd{alberto.rivadulla.sanchez@usc.es}
\emailAdd{davidvazquez.rodriguez@usc.es}

\abstract{We study the particle content of higher derivative theories of gravity built with contractions of the Riemann tensor and its covariant derivatives. In the absence of the latter, there is a family of theories exhibiting an Einsteinian spectrum known as generalized quasi-topological gravities. In turn, we present a no-go result for the construction of Einsteinian gravities involving covariant derivatives of the Riemann tensor. We find evidences suggesting that (truncated series) finite order Lagrangians with covariant derivatives of the Riemann tensor generically present ghosts in their spectrum. This might be interpreted as a hint of non-locality in any healthy UV completion of General Relativity.}

\begin{document}
\maketitle
\flushbottom

\section{Introduction}
\label{sec:Introduction}

Half a century ago, David Lovelock proposed and showed a remarkable theorem \cite{Lovelock}: the most general, symmetric and divergence free tensor which is quasi-linear in the second derivatives of the metric without involving higher derivatives can be obtained from a Lagrangian corresponding to a linear combination of dimensionally continued Euler densities. An immediate corollary of this theorem is that there is only room for the Einstein-Hilbert and cosmological terms in four dimensions; in order to write Lovelock Lagrangians involving polynomials of order $k$ in the curvature, we need to go to spacetime dimensions with $d \geq 2k + 1$.

Allowing the inclusion of higher derivatives in the equations of motion has an old dated bad reputation. The spectrum is modified: extra degrees of freedom appear and ghosts tend to be among them on general grounds. In that respect, Lovelock theories provide a safe playground to explore many interesting aspects of the gravitational interaction, albeit in higher dimensions. Among these, we can include features of black holes such as their existence and uniqueness theorems, their thermodynamics, the definition of their mass and entropy, etc. There are also physically sound problems in higher dimensions in the context of the AdS/CFT correspondence, and Lovelock theory has something to say about them (see, for example, \cite{CESdS}).

If we want to relax the stringent requirements of the Lovelock theorem but stick to the rule that Lagrangian densities are built from contractions of the Riemann curvature and the metric, the equations of motion become fourth order. This does not mean that their study is unworthy. In the late seventies, quadratic terms were scrutinized in view of renormalizing the linearized version of General Relativity (GR) \cite{Stelle1, Stelle2}.

More recently, a new avenue to explore higher curvature corrections to GR was opened after the construction of so-called Einstenian Cubic Gravity (ECG) \cite{ECG}: a cubic theory possessing second-order linearized equations around maximally symmetric backgrounds and admitting Schwarzschild-like solutions characterized by a single function, $g_{tt} g_{rr}=-1$. This was soon extended to a more ample family of theories dubbed Generalized Quasitopological Gravities (GQTGs) \cite{GQTG}; interestingly enough, it was recently shown that all theories involving invariants constructed from contractions of the Riemann tensor and the metric are equivalent via field redefinitions to these \cite{Bueno:2019ltp}, and that GQTGs can be found at all orders \cite{GQTGAllOrders}, thus generalizing the original result \cite{GQTG} where only cubic contractions were considered. Both ECG and GQTGs have been shown to lead to a well-posed cosmological initial-value problem after a suitable choice \cite{Arciniega:2018fxj, Arciniega:2018tnn}, opening an intriguing scenario for cosmological inflation \cite{Edelstein:2020nhg, Edelstein:2020lgv}.

Albeit these bottom-up constructed theories might be afflicted by stability, unitarity or causality issues, it is interesting to explore them for several reasons ranging from phenomenological to theoretical. Among the former we may include cosmological problems such as inflation \cite{Arciniega:2018fxj, Arciniega:2018tnn, Edelstein:2020nhg, Edelstein:2020lgv}, dark energy or the so-called Hubble tension. The latter, instead, have to do with a deeper understanding of General Relativity as an effective field theory and its would be UV completion. It was recently shown, for instance, that quadratic or cubic (in the Riemann tensor) corrections entail causality violation and an ill-behaved eikonal graviton scattering, which suggest the necessity of a stringy UV completion \cite{CEMZ} (see also \cite{Veneziano, EGGKLT, Minwalla1, Minwalla2, EGLS}).

Higher derivative corrections to GR must also include operators involving explicit covariant derivatives of the Riemann tensor. For instance, a $(\nabla Riemann)^2$ term is as valid ---and contributes at the same order in derivatives--- as all $Riemann^3$ contractions in the effective Lagrangian. In the framework of the previous discussion, it is natural to ask ourselves whether we can build Lagrangian densities involving explicit derivatives of the Riemann tensor which preserve some of the nice properties displayed by GQTGs (or their cosmological extensions). We certainly expect extra degrees of freedom arising from the higher derivative equations of motion. However, it is not {\it a priori} obvious whether there exists a choice of the Lagrangian parameters, a family of theories, where the mass of the undesired particles can be taken to infinity thereby removing them from the low energy spectrum, as it is the case in the above referred GQTGs \cite{aspects}.

This is precisely the main objective of this work: to study the particle content when $\nabla Riemann$ terms are considered and scrutinize the possible existence of well behaved Lagrangians of this form. To this end we will start by generalising a well-known construction by Padmanabhan  \cite{Padmanabhan}, valid for $\mathcal{L}(g_{ab}, R \indices{^a _b _c _d})$ theories, to allow for the inclusion of explicit covariant derivatives acting on the Riemann tensor. We will show that all Lagrangians with a finite number of terms of this form generically contain ghosts in the spectrum which cannot be removed. Possible ways to escape this result will also be examined and shown to lead to other problems.

We start by considering in full detail the most general Lagrangian with $(\nabla Riemann )^2$ terms. As we mentioned earlier, these are bound to accompany the $Riemann^3$ terms from the viewpoint of Effective Field Theory; they both scale as $E^6$. The substantial difference between them relies in the (derivative) order of the Euler-Lagrange equations: the former leads to sixth order equations of motion while the latter are fourth order. This very fact implies an extension in the spectrum that must be discussed in detail; ghosts are shown to be unavoidable.

A way-out to this no-go result seems to exist when we embed these Lagrangians into a broader class of theories, all leading to sixth order equations of motion. This can be easily done by writing terms roughly of the form $f(Riemann) (\nabla Riemann)^2$. If we do so, it is possible to write down finite order Lagrangians involving covariant derivatives of the Riemann tensor where the ghosts masses tend to infinity thereby decoupling from the low energy dynamics. However, these involve operators scaling (at least) as $E^8$ thereby demanding the inclusion of all possible terms with such scaling. In particular, this demands the introduction of even higher covariant derivatives of the Riemann tensor and we show that these reintroduce ghost degrees of freedom in the spectrum. Therefore, we argue that this kind of truncation, albeit possible, is unnatural from the point of view of Effective Field Theory.

We finally discuss what are the implications of these results and how they relate to other papers in the literature. In summary, we add further pieces of evidence suggesting that the higher-derivative series cannot be truncated and all operators must be included at once, albeit with special values of the coefficients that come from the ultimate necessity of a physically sensible UV completion.

\section{Derivation of the equations of motion}
\label{sec:Derivation of the equations of motion}

In this section we would like to generalize Padmanabhan's classic construction \cite{Padmanabhan}, which was derived in the context of $\mathcal{L}(g_{ab}, R \indices{^a _b _c _d})$ theories, to a more general family where explicit derivatives acting on curvature tensors appear in the Lagrangian,
\begin{equation}
S = \int d^{d}x \sqrt{|g|}\ \mathcal{L}(g_{ab}, R \indices{^a _b _c _d}, \nabla_{e} R \indices{^a _b _c_d}) ~.
\label{action}
\end{equation}
We will later discuss the more general case involving $\nabla^{2n} Riemann$ terms but, for the moment, we stick to single covariant derivatives acting on the curvature; {\it i.e.}, theories whose equations of motion are sixth order. We treat $g_{ab}$, $R \indices{^a_b_c_d}$ and $\nabla_e R \indices{^a_b_c_d}$ as primary independent fields in order to perform the variation
\begin{equation} 
\delta S = \int d^{d}x \sqrt{|g|}  \Big[ \left( P^{ab} - \frac{\mathcal{L}}{2} g^{ab} \right) \delta g_{ab} + P \indices{_a ^b ^c ^d}\;\delta R \indices{^a _b _c _d} + Q \indices{^e_a^b^c^d }\;\delta \nabla_{e} R \indices{^a _b _c_d} \Big] ~,
\label{deltaS}
\end{equation}
where we have defined the tensors
\begin{equation}
P^{ab} := \frac{\partial \mathcal{L}}{\partial g_{ab}}  ~, \qquad P^{abcd} := \frac{\partial \mathcal{L}}{\partial R_{abcd}} ~, \qquad Q \indices{^e _a^b^c^d } := \frac{\partial \mathcal{L}}{\partial \nabla_{e} R \indices{^a_b_c_d}} ~,
\label{Definition tensor derivatives of Lagrangian}
\end{equation}
and in each variation it is (unless otherwise stated) always assumed that the other primary fields are held constant. The kind of theories we are interested in are those where $Q \indices{^e _a^b^c^d } \neq 0$. The derivatives with respect to the metric and both the Riemann curvature and its covariant derivative are certainly not independent. This can be seen by working out the Lie derivative of $\mathcal{L}$ corresponding to an infinitesimal diffeomorphism, $x^a \to x^a + \xi^a(x)$, which can be alternatively written in two different forms
\begin{equation}
\delta \mathcal{L} = \xi^m \nabla_m \mathcal{L} = P_{ab} \mathcal{L}_\xi g^{ab} + P\indices{_a^b^c^d} \mathcal{L}_\xi R\indices{^a_b_c_d} + Q \indices{^e_a^b^c^d} \mathcal{L}_\xi \nabla_e R\indices{^a_b_c_d} ~;
\label{Lietwoforms}
\end{equation}
the former reads
\begin{equation}
\delta \mathcal{L} = \xi^m \left( P_{ab} \nabla_m g^{ab} + 	P\indices{_a^b^c^d} \nabla_m R\indices{^a_b_c_d} + Q\indices{^e_a^b^c^d} \nabla_m \nabla_e R\indices{^a_b_c_d} \right) ~,
\label{Lieformer}
\end{equation}
while the latter is given by
\begin{eqnarray}
\nonumber \delta \mathcal{L} & = & -2 P_{ab} \nabla^a \xi^b + P\indices{_a^b^c^d} \xi^m \nabla_m R\indices{^a_b_c_d} + 2 P\indices{_{(a|}^c^d^e} R_{{|b)}cde} \nabla^b \xi^a + Q\indices{^e_a^b^c^d} \xi^m \nabla_m \nabla_b R\indices{^a_b_c_d} \\ [0.4em]
& & \hspace{1cm} + Q\indices{_{(a|}_c^d^e^f} \nabla_{|b)} R\indices{^c_d_e_f} \nabla^b \xi^a + 2 Q\indices{_c_{(a|}^d^e^f} \nabla^c R\indices{_{|b)}_d_e_f} \nabla^b \xi^a ~.
\label{Lielatter}
\end{eqnarray}
Comparing both expressions leads to
\begin{equation}
P^{ab} = \frac{\partial \mathcal{L}}{\partial g_{ab}} = \mathcal{R}^{ab} + \frac{1}{2} Q^{{(a|}ijkl}\nabla^{|b)} R_{ijkl} + Q^{l(a|ijk} \nabla_l R \indices{^{|b)} _{ijk}} ~,
\label{identidade1}
\end{equation}
where $\mathcal{R}^{ab} := P^{aijk} R \indices{^b_{ijk}}$, which generalizes Padmanabhan's expression \cite{Padmanabhan} by the inclusion of the novel $Q$-terms. We will see below that these new terms have no effect in the spectrum around a maximally symmetric spacetime (MSS). The last term in \eqref{deltaS} can be written as
\begin{equation}
\begin{array}{ccl}
Q \indices{_{ea}^{bcd}}\delta \nabla^e  R \indices{^a _b_c_d} & = & Q \indices{_{ea}^{bcd}} [\delta,\nabla^e ]R \indices{^a _b_c_d} + Q \indices{_{ea}^{bcd}} \nabla^e \delta R \indices{^a _b_c_d} \\ [0.9em]
& = & Q \indices{_{ea}^{bcd}} [\delta,\nabla^e] R \indices{^a _b_c_d} - \nabla^e \left(Q \indices{_{ea}^{bcd}} \right) \delta R \indices{^a _b_c_d} + \nabla^e \left( Q\indices{_e_a^{bcd}} \delta R\indices{^a_{bcd}} \right)~,
\end{array}
\label{Q*commutator}
\end{equation}
after integrating by parts. Now, the commutator between the variation and the covariant derivative acting on the Riemann tensor reads
\begin{equation}
[\delta,\nabla_e] R \indices{^a _b_c_d} = \delta \Gamma \indices{^a _e _f} R \indices{^f _b_c_d}-\delta \Gamma \indices{^f _e _b} R \indices{^a _f_c_d}- \delta \Gamma \indices{^f _e_c} R \indices{^a _b_f_d}- \delta \Gamma \indices{^f _e _d} R \indices{^a _b_c_f} ~,
\label{commutatordeltanabla}
\end{equation}
where
\begin{equation} 
\delta \Gamma \indices{^a_b_c} = \frac{1}{2} g^{ad} (- \delta g_{bc;d} + \delta g_{cd;b} + \delta g_{bd;c}) ~.
\label{deltaGamma}
\end{equation}
Introducing \eqref{commutatordeltanabla} and the Palatini identity
\begin{equation}
\delta R \indices{^a_b_c_d} = \nabla_c \delta \Gamma \indices{^a_b_d} - (c \leftrightarrow d)
\label{palatini}
\end{equation}
in the second line of \eqref{Q*commutator}, we have the whole expression written in terms of $\delta \Gamma \indices{^a _b_c}$. Plugging these expressions into \eqref{deltaS} and using the relation between $\delta \Gamma \indices{^a_{bc}}$ and $\delta g_{ab}$ with a proper integration by parts, we can finally obtain the equations of motion,
\begin{equation}
\mathcal{E}_{ab} = \mathcal{E}^{0}_{ab} + \mathcal{E}^{\nabla}_{ab} = \frac{1}{2}T_{ab} ~,
\label{eom}
\end{equation}
where $\mathcal{E}^{0}_{ab}$ contains the well-known terms involving no covariant derivatives of the Riemann tensor \cite{Padmanabhan,ECG,aspects},
\begin{equation}	
\mathcal{E}_0^{ a b} = \mathcal{R}^{ab} - \frac{1}{2} g^{ab} \mathcal{L} - 2 \nabla_d \nabla_c P \indices{^{acdb}} ~,
\label{eom-0}
\end{equation}
while $\mathcal{E}^{\nabla}_{ab}$ is the new contribution precisely coming from those covariant derivatives and the extension on $P^{ab}$ proportional to $Q \indices{_e _a ^b ^c ^d}$ \eqref{identidade1},
\begin{equation} 
\begin{array}{ccl}
\mathcal{E}_{\nabla}^{ab} & = & 2 \nabla_d \nabla_c \nabla_e Q \indices{^e^a ^c ^d ^b} -  \nabla_c \Big(2 Q \indices{^{ab}_{ijk}}R^{icjk}+2Q\indices{^{acijk}} R\indices{^b _{ijk}}+Q \indices{^{caijk}} R\indices{^b _{ijk}} \Big) \\ [0.9em]
& & \hspace{1.3cm} + \frac{1}{2} Q^{a ijkl}\nabla^{b} R_{ijkl} + Q^{l a ijk} \nabla_l R \indices{^{b}_{ijk}} ~.
\end{array}
\label{eom-nabla}
\end{equation}
$T_{ab}$ is the energy-momentum tensor of the matter fields that could be added to the gravitational action \eqref{action}. Notice that in GR, $P_{abcd}= \frac{1}{2 \kappa} \left( g_{ac} g_{bd} - g_{ad} g_{bc} \right)$ and $Q_{eabcd}$ is zero, so $\mathcal{R}_{ab}$ is simply the Ricci tensor and we recover Einstein's equations.

Now, in order to study the spectrum of the theory, we can generalize the method developed in \cite{ECG, aspects} to account for the inclusion of $\nabla Riemann$ terms. Varying the action \eqref{action} and keeping track of the boundary terms we find
\begin{equation}
\delta S = \int_{\mathcal{M}} d^d x \sqrt{-g} \, \mathcal{E}_{ab} \delta g ^{ab} + \int_{\mathcal{M}} d^{d-1}x \sqrt{-g} \, \nabla_a \delta v^a ~,
\label{varactionbdy}
\end{equation}
where
\begin{equation}
\begin{array}{ccl}
\delta v^c & = & \delta v_0^c
+ 2 Q\indices{_a^{cdef}} R_{bdef} \delta g^{ab}
+ Q\indices{^c_a^{def}} R_{bdef} \delta g^{ab}
- 2 Q\indices{_{ab}^{def}} R\indices{^c_{def}} \delta g^{ab} \\ [0.9em]
& & \hspace{0.5cm} -2 Q\indices{^{cfbed}} \nabla_e \nabla_b \delta g_{fd}
- 2 \nabla^e Q\indices{_e^{abcd}} \nabla_a \delta g_{bd}
+ 2 \nabla_a \nabla^e Q\indices{_e^{cbad}} \delta g_{bd} ~,
\end{array}
\end{equation}
and $\delta v_0^c = 2 P^{abcd} \nabla_b \delta g_{da} - 2 \delta g_{da} \nabla_b P^{acbd}$ is the $\mathcal{L}(Riemann)$'s boundary contribution  already studied in \cite{Padmanabhan}. Since we will consider linearized perturbations in (A)dS spacetime, we fix the transverse gauge,\footnote{Even though our results were derived considering an (A)dS background, the procedure to find the spectrum can be extended to flat space ---and the graviton masses \eqref{gravitons masses} turn out to be the same---, the only subtlety being the necessity of using the de Donder gauge, $\partial_a h^{ab} = \frac{1}{2} \partial^b h$, instead of the transverse gauge in order to be consistent; see, for example, \cite{aspects}.}
\begin{equation}
\label{transversegauge}
\nabla_a h^{ab} = \nabla^b h ~,
\end{equation}
where we adopted the notation $h^{ab} = \delta g^{ab}$ (and, of course, $h = h^a_{\ a}$). Let us now discuss how the linearization procedure applies both to $\mathcal{L}(Riemann)$ and $\mathcal{L}(Riemann, \nabla Riemann)$ contributions.

\subsection{Linearization of the $\mathcal{L}(Riemann)$ contribution}
\label{sec:Linearization of  L(Riemann)}

In this subsection we review the method introduced in \cite{ECG,aspects} to linearize the $\mathcal{L}(Riemann)$ part; {\it i.e.}, all contributions not involving $\nabla Riemann$ terms. We introduce the Riemann-like tensor $r_{abcd}$,
\begin{equation}
r_{abcd} = \bar{R}_{abcd} + 2 \alpha k_{a[c|} k_{|d]b} ~,
\label{pablos method}
\end{equation}
where
\begin{equation}
    \bar{R}_{abcd} = 2 \Lambda\,g_{a[c|} g_{|d]b}
    \label{riemann MSS}
\end{equation}
is the Riemann tensor of the MSS background with cosmological constant $\Lambda$, and we defined the (symmetric) projector $k_{ab}$, verifying $k_{ac} k^{cb} = k \indices{_a ^b}$, and $\chi := k \indices{^a _a}$ is an arbitrary integer constant smaller than $d$. We then evaluate the Lagrangian replacing the Riemann tensor by $r_{abcd}$, such that $\mathcal{L}(r_{abcd}) = \mathcal{L}(\alpha, \chi, \Lambda)$. The linearized equations of motion read
\begin{eqnarray}
\mathcal{E}^{0,L}_{ab} & = & 2 \left[ e - 2\Lambda ((d-1) a + c) + (2a + c) \bar\Box \right] G_{\langle ab \rangle}^{L} + 2 (a + 2b + c) \left[ \bar g_{ab}\,\bar\Box - \bar\nabla_{a} \bar\nabla_{b} \right] R^L \nonumber \\ [0.5em]
& & - 2\Lambda \left[ (d-3) a - 2 (d-1) b - c \right] g_{ab}\, R^L = \frac12 T_{ab}^L ~,
\label{EOL}
\end{eqnarray}
where we used the expressions for the Einstein/Ricci tensors and the scalar curvature at the linearized level
\begin{eqnarray}
G_{ab}^{L} & = & R_{ab}^{L} - \frac12 \bar g_{ab} R^L - (d-1) \Lambda h_{ab} ~, \\ [0.7em]
R_{ab}^{L} & = & \bar\nabla_{(a|} \bar\nabla_{c} h^c_{\ |b)} - \frac12 \bar\Box h_{ab} - \frac12 \bar\nabla_a \bar\nabla_b h + d \Lambda h_{ab} - \Lambda \bar g_{ab} h ~, \\ [0.7em]
R^L & = & \bar\nabla^a \bar\nabla^b h_{ab} - \bar\Box h - (d-1) \Lambda h ~.
\end{eqnarray}
The parameters $a$, $b$, $c$ and $e$ can be computed for any theory as
\begin{eqnarray*}
& \mathit{a} & = \left[ \frac{1}{4 \chi(\chi-1) } \frac{\partial^2 \mathcal{L}}{\partial \alpha^2} \Big|_{\alpha = 0} \right]_{\chi = 1} \! , \qquad\ \mathit{c} = \left[ \frac{1}{\chi-1} \left( \frac{1}{4 \chi(\chi-1) } \frac{\partial^2 \mathcal{L}}{\partial \alpha^2} \Big|_{\alpha = 0} - a \right) \right]_{\chi = 0} ~, \\ [0.9em]
& \mathit{b} & = \frac{1}{\chi(\chi-1)} \left[ \frac{1}{4 \chi(\chi-1) } \frac{\partial^2 \mathcal{L}}{\partial \alpha^2} \Big|_{\alpha = 0} - a - c (\chi-1) \right] ~,  \qquad \mathit{e} = \frac{1}{2 \chi(\chi-1) } \frac{\partial \mathcal{L}}{\partial \alpha} \Big|_{\alpha = 0} ~.
\end{eqnarray*}
Considering perturbations over a background metric $\bar{g}_{ab}$, $g_{ab} = \bar{g}_{ab} + h_{ab}$, we can decompose the field equations into their traceless and trace parts, 
\begin{eqnarray} \label{Epadmanabhan linealized traceless}
\mathcal{E}^{0,L}_{\langle ab\rangle} & = & \displaystyle\frac{1}{4m_g^2\kappa_{\rm eff}} \Bigg\{ \left[\bar\Box -2\Lambda\right]\left[\bar\Box-2\Lambda-m_g^2\right]h_{\langle ab\rangle}-\bar\nabla_{\langle a }\bar\nabla_{b\rangle}\bar\Box h \\ \nonumber 
& & \hspace{17mm} + \displaystyle\left[\frac{m_g^2(m_s^2+2(d-1)\Lambda)+\Lambda((4-3d)m_s^2-4(d-1)^2\Lambda)}{(m_s^2+d\Lambda)} \right]\bar\nabla_{\langle a}\bar\nabla_{b\rangle}h \Bigg\} ~, \\[0.4em]
\label{Epadmanabhan trace}
\mathcal{E}^{0,L} & = & - \displaystyle\left[\frac{(d-1)(d-2)\Lambda(m_g^2-(d-2)\Lambda)}{4\kappa_{\rm eff}m_g^2(m_s^2+d\Lambda)}\right]\left[\bar\Box -m_s^2\right]h ~,
\end{eqnarray}
where we have chosen a different set of (physically more meaningful) parameters, $\kappa_{\text{eff}}$, $m_g^2$ and $m_s^2$, which represent respectively the effective Newton constant (modified due to the higher curvature terms) and the masses of the ghost graviton and the scalar field customarily appearing as extra degrees of freedom in a general higher curvature theory of the type $\mathcal{L}(Riemann)$ \cite{aspects}. The relations between both sets are:
\begin{eqnarray*}
\mathit{a} & = & \frac{1}{4\Lambda(d-3)\kappa_{\rm eff}} \left( 2 e \kappa_{\rm eff} - \frac12 \right) ~, \\ [0.7em]
\mathit{b} & = & \frac{1}{4\Lambda(d-3)\kappa_{\rm eff}} \left[ \frac{(4 e \kappa_{\rm eff}-1)(d-1)m_s^2 +2(3-2d+2(d-1)de \kappa_{\rm eff}) \Lambda}{4 (d-1)(m_s^2+d\Lambda)} \right. \\
& & \left. \hspace{67mm} + \frac{(d-3)\Lambda(d m_s^2+4(d-1)\Lambda)}{4m_g^2(d-1)(m_s^2+d\Lambda)} \right] ~, \\ [0.7em]
\mathit{c} & = & \frac{1}{4\Lambda(d-3)\kappa_{\rm eff}} \left( 1 - 4e \kappa_{\rm eff} - \frac{(d-3) \Lambda}{m_g^2} \right) ~, \\
\\
\mathit{e}&=& \frac{ 4 (d-1) \Lambda}{\bar{\mathcal{L}}} ~,
\end{eqnarray*} 
where $\bar{\mathcal{L}}$ is the Lagrangian evaluated on the MSS background \eqref{riemann MSS}.
Having reviewed the procedure along the lines developed in \cite{ECG,aspects}, we are now ready to explore the general case involving covariant derivatives of the Riemann tensor.

\subsection{Linearization of the $\mathcal{L}(Riemann, \nabla Riemann)$ contribution}
\label{sec-Linearization of the L(Riemann, nabla Riemann) contribution}

First of all, let us discuss the kind of terms that can appear in a general Lagrangian of the form $\mathcal{L}(Riemann, \nabla Riemann)$. Notice that the Riemann/Ricci tensors and the scalar curvature are all objects with an even number of indices. This implies that we will need to have always an even number of explicit covariant derivatives contracted with these objects in order to obtain scalar quantities. That is to say, and in a symbolic manner, all possible contractions have to be of the form $(Riemann)^k (\nabla Riemann)^{2j}$, with integers $k, j \geq 0$. The discussion carried out in this subsection takes this into account.

Let us first discuss why terms involving $\nabla Riemann$ in the Lagrangian do not contribute to the linearization of $\mathcal{E}^{0}_{ab}$. Recall that its general expression \eqref{eom-0} contains the objects $\mathcal{L}$ and $P^{abcd}$, whose variation yielded the terms obtained above. Now, for a Lagrangian of the form $\mathcal{L}(Riemann, \nabla Riemann)$, some new contributions arise since
\begin{equation}
\delta \mathcal{L} = \bar{P}_{ab}\,\delta g^{ab} + \bar{P}\indices{_a^b^c^d}\,\delta R\indices{^a_b_c_d} + \bar{ Q}\indices{^e_a^b^c^d}\,\delta\nabla_e R\indices{^a_b_c_d} ~,
\end{equation}
and
\begin{equation}
\delta P^{abcd} = \frac{\partial\bar{P}^{abcd}}{\partial g^{ef}} \delta g^{ef}+ \frac{\partial\bar{P}^{abcd}}{\partial R\indices{^e_{fgh}}} \delta R\indices{^e_{fgh}} + \frac{\partial \bar{P}^{abcd}}{\partial \nabla_e R\indices{^f_g_h_i}} \delta \nabla_e R\indices{^f_g_h_i} ~,
\label{deltaPabcd}
\end{equation}
where all terms of the form $\partial\bar{P}^{abcd}/\partial\Psi$ must be understood as the variation with respect to $\Psi$ followed by the evaluation in the background MSS whose Riemann tensor is given by \eqref{riemann MSS}. Recall that even the tensor $P^{ab}$ is affected by the existence of $\nabla Riemann$ terms in the Lagrangian, its full expression written in \eqref{identidade1}. These terms are restricted, as explained above, to be of the form $(Riemann)^k (\nabla Riemann)^{2j}$, which implies that $Q\indices{_e^a_b_c_d} \sim (Riem)^k (\nabla Riem)^{2j-1}$, and thereby $\bar{Q}\indices{_e^a_b_c_d}$ vanishes on a MSS. The last term in \eqref{deltaPabcd} is proportional to $(Riemann)^{k-1} (\nabla Riemann)^{2j-1}$, so it also vanishes when evaluated on this background. The additional terms in $P^{ab}$ \eqref{identidade1} contain contractions of the form $Q\, \nabla Riemann$, which do not contribute when evaluated on the background, since $\nabla Riemann = 0$ in this case. Finally, notice that
$$
\frac{\partial P^{abcd}}{\partial g^{ef}} = \left. \frac{\partial P^{abcd}}{\partial g^{ef}} \right|_{\nabla Riemann = 0} + \mathcal{O}\left( (\nabla Riemann)^{2j} \right) ~,
$$
the first term corresponding to the value of this quantity computed only from the part of the Lagrangian without explicit covariant derivatives of the Riemann tensor, and the last term vanishing in the background. In summary, the linearized equations obtained from $\mathcal{E}_0^{ab}$ do not receive new contributions coming from terms in the Lagrangian that involve contractions of $\nabla Riemann$.

We shall now consider the linearization of $\mathcal{E}_\nabla^{ab}$, given in \eqref{eom-nabla}, in order to understand the implications in the general spectrum of the terms containing explicit derivatives of the Riemann tensor in the Lagrangian. We will need the variation
\begin{equation} \label{deltaQ}
\delta Q^{abcde}= \frac{\partial \bar{Q}^{abcde}}{\partial g_{ij}} \delta g_{ij}+  \frac{\partial \bar{Q}^{abcde}}{\partial R_{ijkl}} \delta R_{ijkl} + \bar{W} \indices{^{abcde} _{hijkl}} \delta \nabla^h R^{ijkl} ~,
\end{equation}
where, again, all terms of the form $\partial\bar{Q}^{abcde}/\partial\Psi$ must be understood as variations followed by evaluation in the background \eqref{riemann MSS}, and we have defined
$$
W \indices{^{abcde}_{hijkl}} := \frac{\partial Q^{abcde}}{\partial \nabla^h R^{ijkl}} ~.
$$
Taking into account the arguments presented at the beginning of this section about the form of the contractions that can appear in the Lagrangian, we can make a couple of comments about the general form of $Q^{abcde}$. First of all, recall that $Q^{abcde}$ itself must vanish on a MSS vacuum. In the case of $\delta Q^{abcde}$, much for the same reason, there are neither contributions proportional to $\delta g_{ij}$ nor to $\delta R_{ijkl}$, since both $\frac{\partial \bar{Q}^{abcde}}{\partial g_{ij}}$ and $\frac{\partial \bar{Q}^{abcde}}{\partial R_{ijkl}}$ are zero.

We conclude that the only possible contribution to $\delta Q^{abcde}$ at linear level comes from terms proportional to the second derivative of the Lagrangian with respect to $\nabla Riemann$,
\begin{equation} \label{delta Q in m.s.s}
\delta Q^{abcde}= \bar{W} \indices{^{abcde}_{hijkl}}\, \delta\nabla^{h} R\indices{^{ijkl}} ~.
\end{equation}
Therefore, any term of the form $(\nabla Riemann)^{2j} (Riemann)^k$ with $j \in \mathbb{Z}^+$, $j>1$  \textit{does not modify the linearized spectrum}. Any such term would leave a remnant proportional to $\nabla Riemann$ in $\bar{W} \indices{^{abcde}_{hijkl}}$, which vanishes when evaluated on a MSS.\footnote{This provides an infinite set of terms with an Einstein-spectrum on a MSS background, something that might have interesting applications in the realm of the AdS/CFT correspondence, given that we are led to a plethora of gravity actions and black holes whose asymptotic behavior is free from ghosts and extra degrees of freedom living at the boundary.} For this reason, in the next section we will stick to the case $j=1$,
$$
\mathcal{L} = \mathcal{L}(Riemann) + \mathcal{L}_{\nabla}(Riemann, (\nabla Riemann)^2) ~,
$$
whose linearized equations of motion are $\mathcal{E}_L^{ab} = \mathcal{E}_{0, L}^{ab} + \mathcal{E}_{\nabla, L}^{ab}$, where the first part is due entirely to $\mathcal{L} (Riemann)$ as was argued before ---and studied in section \ref{sec:Linearization of  L(Riemann)}---, while $\mathcal{E}_{\nabla, L}^{ab}$ is the novel contribution produced by the terms with contractions of $\nabla Riemann$, which in this case is found to be
\begin{eqnarray}
\mathcal{E}_{\nabla,L}^{ab} & = & 4 \Big[  \Lambda^2 \bar{W} \indices{_f ^{agb} _{gec} ^{h}_{dh}} \bar{\nabla}^f \bar{\nabla}^e + \Lambda \Big( \bar{W} \indices{_h ^{aib}_{igcedf}} + \bar{W} \indices{_f^a_g^b_{hec}^i_{di}} \Big) \bar{\nabla}^h \bar{\nabla}^g \bar{\nabla}^f \bar{\nabla}^e \nonumber \\ [0.7em]
& & \hspace{13mm} + \bar{W} \indices{_h^a _i ^b _{jgcedf}} \bar{\nabla}^j \bar{\nabla}^i \bar{\nabla}^h \bar{\nabla}^g \bar{\nabla}^f \bar{\nabla}^e \Big] h^{cd} ~.
\label{eom nabla lineal part}
\end{eqnarray}
%
In order to study a particular theory we just need to compute the form of the tensor $W \indices{^{abcde}_{hijkl}}$, evaluate it on a MSS, and read the spectrum using \eqref{eom nabla lineal part}.

\section{Spectrum of the $(\nabla Riemann)^2$ theory} 

Considering all possible contractions of the form $(\nabla Riemann)^2$, the most general Lagrangian (to lowest order in the higher derivative expansion) would be:
\begin{equation}
\mathcal{L}_{\nabla} = \alpha\, \nabla_b R_{ac} \nabla^c R^{ab} + \beta\, \nabla_e R_{abcd} \nabla^e R^{abcd} + \mu\, \nabla_a R \nabla^a R + \nu\, \nabla_a R_{bc} \nabla^a R^{bc} ~,
\label{nabla lagrangian}
\end{equation}
where $\alpha, \beta, \mu, \nu$ are coupling constants with energy dimension $E^{d-6}$. However, if we use Bianchi identities and integrations by parts, we can readily see that
$$
\mathcal{L}_{\alpha} := \alpha\, \nabla_b R_{ac} \nabla^c R^{ab} = \frac{\alpha}{4} \nabla_a R \nabla^a R + \mathcal{O}(Riemann^3) + \text{total derivative} ~.
$$
This means that $\mathcal{L}_{\alpha}$ is redundant in our analysis, and can be taken away by suitably modifying the parameter $\mu$, as well as those corresponding to the cubic terms in $\mathcal{L}(Riemann)$ ($a$, $b$, $c$ and $e$, defined above). Something similar happens with $\mathcal{L}_{\beta} := \beta\, \nabla_e R_{abcd} \nabla^e R^{abcd}$, which can be reshuffled as
$$
\mathcal{L}_{\beta} = - \beta\, \nabla_a R \nabla^a R + 4 \beta\, \nabla_a R_{bc} \nabla^a R^{bc} + \mathcal{O}(Riemann^3) + \text{total derivative} ~,
$$
after integrating by parts and using the Bianchi identities. Again, $\mathcal{L}_{\beta}$ is shown to be redundant in our analysis, and can be removed at the expense of modifying the parameters $\mu$, $\nu$, $a$, $b$, $c$ and $e$. To sum up, we could (and will) use \eqref{nabla lagrangian} with $\alpha = \beta = 0$,
\begin{equation}
\mathcal{L}_{\nabla} = \mu\, \nabla_a R \nabla^a R + \nu\, \nabla_a R_{bc} \nabla^a R^{bc} ~,
\label{nabla lagrangian final}
\end{equation}
together with the $\mathcal{L}(Riemann)$ contribution, where $\mu$ and $\nu$ (as well as those parameters related to the terms with no explicit covariant derivatives of the Riemann tensor) have been redefined in order to conveniently reabsorb $\alpha$ and $\beta$. This is relevant when comparing our results with other works in the literature where an explicit dependence on $\alpha$ and $\beta$ is considered.

\subsection{Linearization of the field equations}
\label{subsec-Linearization Lagrangian (Nabla Riemann)2}

There is another observation to make before moving on to the general discussion of the spectrum. Notice that the insertion of \eqref{nabla lagrangian final} does not modify the cosmological constant $\Lambda$ on MSS solutions. This can be easily seen, since both $\bar{\mathcal{L}}_{\nabla}$ and $\bar{\delta} \mathcal{L}_{\nabla}$ vanish. Therefore, $\bar{\mathcal{E}}_{\nabla}^{ab} = 0$; these terms do not contribute to the vacuum equations of motion.

Considering the transverse gauge \eqref{transversegauge}, the (traceless and trace) linearized equations of motion, $\mathcal{E}_{L}^{ab} := \mathcal{E}_{0,L}^{ab} + \mathcal{E}_{\nabla, L}^{ab}$, read:
\begin{eqnarray}
\mathcal{E}_{L}^{\langle ab \rangle} & = & \hat{\kappa}^{-1} \Big[ (\bar{\Box} - 2 \Lambda) (\bar{\Box} - 2 \Lambda - m_{\xi_+}^2) (\bar{\Box} - 2 \Lambda - m_{\xi_-}^2)\, h^{\langle a b \rangle} \qquad \nonumber \\ [0.5em]
& & \hspace{13mm} \qquad - (\bar{\Box} - c_1) (\bar{\Box} - c_2) \bar{\nabla}^{\langle a \,} \bar{\nabla}^{\,  b\rangle}\, h  \Big] = \frac{1}{2} T_{L}^{\langle ab \rangle} ~, \label{eom traceless total ads} \\ [0.7em]
\mathcal{E}_{L} & = & \kappa^{*\, -1} (\bar{\Box} - m_{\phi_+}^2) (\bar{\Box} - m_{\phi_-}^2)\, h = \frac{1}{2} T_L ~,
\label{trace total ads}
\end{eqnarray}
where we have defined the quantities
\begin{equation} 
\hat{\kappa}^{-1} = \frac{\nu}{2} ~, \qquad {\rm and} \qquad \kappa^{*\, -1} = \frac{1}{2} (d-1) \Big[ 4 (d-1) \mu + d \nu \Big] \Lambda ~, 
\label{kappas}
\end{equation}
and the (squared) masses
\begin{eqnarray} \label{gravitons masses}
m_{{\xi}_\pm}^2 & = & m_{\xi}^2 \left( 1 \pm  \sqrt{1 - \Delta_\xi} \right) ~, \\ [0.7em]
m_{{\phi}_\pm}^2 & = & m_{\phi}^2 \left( 1 \pm  \sqrt{1 - \Delta_\phi} \right) ~,
\label{scalar masses}
\end{eqnarray}
where $m_{\xi}^2$ and $m_{\phi}^2$,
\begin{eqnarray} \label{mass-g}
m_{\xi}^2 & = & \frac{2 \mathit{a} + \mathit{c} - \nu \Lambda}{\nu} ~, \\ [0.7em]
m_{\phi}^2 & = & \frac{2 \mathit{a} + d \mathit{c} -(d-1)(-4 \mathit{b} + 2 d \Lambda \mu) - (d (d-2) + 2) \Lambda \nu}{4 (d-1) \mu + d \nu} ~,
\label{mass-s}
\end{eqnarray}
are the values of the graviton and scalar masses in theories with vanishing discriminants, $\Delta_\xi = \Delta_\phi = 1$,
\begin{eqnarray} \label{graviton-disc}
\Delta_\xi & = & \frac{2 \nu (2 \mathit{a} (d-3) \Lambda - \mathit{e})}{(2 \mathit{a} + \mathit{c} - \nu \Lambda)^2} ~, \\ [0.7em]
\Delta_\phi & = & - \frac{2 (4 (d-1) \mu +d \nu ) (4 \Lambda  (\mathit{a}+(d-1) (\mathit{c}+\mathit{b} d))+(2-d) \mathit{e})}{(2 \mathit{a} +d \mathit{c} -(d-1)(-4 \mathit{b}+2d \Lambda \mu)-(d(d-2)+2)\Lambda \nu)^2} ~.
\label{scalar-disc}
\end{eqnarray}
$c_1$ and $c_2$ in \eqref{eom traceless total ads} are combinations of the coupling constants, $\Lambda$ and $d$, whose exact form is irrelevant since that part of the equation of motion will be decoupled with a field redefinition in the next section.

A few comments are in order. First of all, notice that the limit $\mu, \nu \to 0$ is tricky: several of the above quantities seem to diverge in such limit, but this simply reflects the fact that it has to be taken at the level of the equations of motion \eqref{eom traceless total ads} and \eqref{trace total ads}. In fact, it is easy to see that $\hat{\kappa}^{-1}, \kappa^{*\, -1} \to 0$ in that limit, and the masses diverge accordingly, in such a way that the terms with maximum number of boxes, $\bar{\Box}$, vanish; the spectrum reducing to that studied in \cite{aspects}.

On general grounds, we will show that the spectrum possesses two massive graviton modes, $\xi_\pm$, and two massive scalar modes, $\phi_\pm$, besides the massless graviton. At this stage, though, we can already see that the massive graviton sector is insensitive to the gravitational coupling $\mu$. In order to discuss the spectrum of the theory we have to decouple the trace degree of freedom, $h$, from the traceless part, $h_{\langle a b \rangle}$, in \eqref{eom traceless total ads}. We can do that by a suitable field redefinition,
\begin{equation} \label{redefinition fields}
t_{ab} := h_{\langle a b \rangle} + \bar{\nabla}_{\langle a} \bar{\nabla}_{b \rangle} (\lambda_1 + \lambda_2\,\bar{\Box}) h ~,
\end{equation}
where $\lambda_1$ and $\lambda_2$ are (soon to be fixed) free constants and $t_{ab}$ is traceless by construction. Plugging \eqref{redefinition fields} into \eqref{eom traceless total ads}, we will have  a `remnant' $h$--dependent part involving terms of the form\footnote{We commuted covariant derivatives using the fact that $h$ is a linearized mode on a MSS.} $\bar{\nabla}_{\langle a} \bar{\nabla}_{b \rangle}\,\bar{\Box}^k h$, with $k = 0, \dots, 4$. Now we can use the trace equation \eqref{trace total ads} in order to replace every $\bar{\Box}^2 h$ in terms of $\bar{\Box} h$, $h$ and $T^L$,
$$
\bar{\Box}^2 h = \frac{\kappa^*}{2} T^L + (m_{\phi_+}^2 + m_{\phi_-}^2) \bar{\Box} h + m_{\phi_+}^2 m_{\phi_-}^2 h ~,
$$
and iterate as many times as we need to rewrite \eqref{eom traceless total ads} as follows:
$$
(\bar{\Box} - 2 \Lambda) (\bar{\Box} - 2 \Lambda - m_{\xi_+}^2) (\bar{\Box} - 2 \Lambda - m_{\xi_-}^2) t_{ab} + f_1 \bar{\nabla}_{\langle a} \bar{\nabla}_{b \rangle} h + f_2 \bar{\nabla}_{\langle a} \bar{\nabla}_{b \rangle} \bar{\Box} h = \frac{\hat{\kappa}}{2} T_{\langle a b \rangle}^{L,\text{eff}} ~,
$$
where $f_1$ and $f_2$ are algebraic equations involving linearly $\lambda_1$, $\lambda_2$, and complicated expressions of $\Lambda$, $m_{\xi_\pm}^2$, $d$, $\mu$ and $\nu$. It is always possible by construction to (uniquely) choose $\lambda_1$ and $\lambda_2$ such that $f_1 = f_2 = 0$. We are left with a decoupled equation of motion for the (redefined) traceless degrees of freedom,
\begin{equation}
\hat{\kappa}^{-1} (\bar{\Box} - 2 \Lambda) (\bar{\Box} - 2 \Lambda - m_{\xi_+}^2) (\bar{\Box} - 2 \Lambda - m_{\xi_-}^2) t_{ab} = \frac{1}{2} T_{\langle a b \rangle}^{L,\text{eff}} ~.
\label{decoupled equation}
\end{equation}
The price to pay when writing the dynamics in this way is a more involved matter coupling, which we capture writing $T_{\langle a b \rangle}^{L, \text{eff}}$; something similar happens in purely $\mathcal{L}(Riemann)$ theories \cite{aspects}. Now, with the relevant degrees of freedom of the traceless and trace part totally decoupled, we are ready to study the spectrum.

\subsubsection{Traceless modes}
\label{subsec-Traceless modes}

To scrutinize the graviton spectrum we should decompose the $t_{ab}$ tensor in three different modes: the three propagating graviton degrees of freedom of our theory. This decomposition has some ambiguities and pitfalls whose analysis is left to an Appendix. We decompose the traceless equation of motion \eqref{decoupled equation} by defining the following three tensors:
\begin{eqnarray} 
\label{zeta} \zeta_{ab} & := & \alpha_1 ( \bar{\Box} - 2 \Lambda - m_{\xi_+}^2) (\bar{\Box} - 2 \Lambda - m_{\xi_-}^2)\, t_{ab} ~, \\ [0.9em]
\xi^{\pm}_{ab} & := & \alpha_\pm ( \bar{\Box} - 2 \Lambda) (\bar{\Box} - 2 \Lambda - m_{\xi_\mp}^2)\, t_{ab} ~, 
\label{xi-pm} 
\end{eqnarray}
where $\alpha_1$ and $\alpha_\pm$ are free parameters to be fixed by imposing the condition:\footnote{As discussed in Appendix \ref{sec-Appendix}, this expression must be understood as a direct sum with no prefactors.}
\begin{equation} \label{t=zeta+xi+chi}
t_{ab} := \zeta_{ab} + \xi^+_{ab} + \xi^-_{ab} ~.
\end{equation}
Indeed, plugging \eqref{zeta} and \eqref{xi-pm} into \eqref{t=zeta+xi+chi}, we obtain
\begin{equation}
\alpha_1^{-1} = m_{\xi_+}^2 m_{\xi_-}^2 ~, \qquad \text{and} \qquad \alpha_\pm^{-1} = m_{\xi_\pm}^2 \left( m_{\xi_\mp}^2 - m_{\xi_\pm}^2 \right) ~.
\end{equation}
The above defined tensor modes, $\zeta_{ab}$ and $\xi^{\pm}_{ab}$, obey the following equations:
\begin{eqnarray}
- \frac{1}{2 \kappa_{\text{eff}}} (\bar{\Box} - 2 \Lambda) \zeta_{ab} & = & T_{\langle a b \rangle}^{L, \text{eff}} ~, \label{zeta graviton eom} \\ [0.9em]
\frac{1}{2 \kappa_{\xi_\pm}} ( \bar{\Box} - 2 \Lambda - m_{\xi_\pm}^2) \xi^\pm_{ab} & = & \frac{1}{2} T_{\langle a b \rangle}^{L, \text{eff}} ~,
\label{xi graviton eom}
\end{eqnarray}
where $\kappa_{\text{eff}} = - \frac14 \alpha_1 \hat{\kappa}$ and $\kappa_{\xi_\pm} = \frac12 \alpha_\pm \hat{\kappa}$. In particular, the effective Newton constants $\kappa_{\xi_\pm}$ can be written as
\begin{equation}
\frac{1}{2 \kappa_{\xi_\pm}} = - \frac{1}{4 \kappa_{\text{eff}}} \frac{(1 + x)^2}{x} \frac{\Lambda}{m_g^2} \sqrt{1 + \frac{2x}{(x+1)^2} \frac{m_g^2}{\Lambda}} \left( \mp 1 - \sqrt{1 + \frac{2x}{(x+1)^2} \frac{m_g^2}{\Lambda}} \right) ~, \label{keff xi} 
\end{equation}
where we introduced the dimensionless number $x := 4 \Lambda\nu\kappa _{\text{eff}}\, m_g^2$, which weights the relative importance of the $\nabla Riemann$ terms.

Let us briefly end by showing how to recover the (already known) results of $\mathcal{L}(Riemann)$ theories \cite{aspects}. First, notice that the coupling $\mu$ is absent from all our expressions; this is due to the fact that the term $\nabla_a R\,\nabla^a R$ only affects the scalar (trace) degrees of freedom. The limit $\nu \to 0$ ($x \to 0$) is tricky, as stated above, since some quantities diverge: $1/\kappa_{\xi_+} \footnote{ In the case where we consider $\mathcal{L}(Riemann)$ theories with an Einsteinian spectrum, \textit{i.e.} $m_g^2 \rightarrow \infty$, this quantity is not divergent and tends to the usual value $1/\kappa_\text{eff}$. However, the mass still diverges, $m_{\xi_+}^2 \rightarrow \infty$, so the mode $\xi_+$ disappears regardless of this.}, m^2_{\xi_+} \to \infty$. This not only entails the full decoupling of the mode $\xi_+$ but its complete disappearance from the spectrum of the theory. Meanwhile,
\begin{eqnarray}
- \frac{1}{2 \kappa_{\text{eff}}} (\bar\Box - 2 \Lambda) \zeta_{ab} & = T_{\langle a b \rangle}^{L,\text{eff}} ~, \qquad & \zeta_{ab} \; \longrightarrow \; \text{Einstein graviton,} \label{zeta graviton limit} \\ [0.9em]
+ \frac{1}{2 \kappa_{\text{eff}}} (\bar\Box - 2 \Lambda - m_g^2) \xi^-_{ab} & = T_{\langle a b \rangle}^{L,\text{eff}} ~, \qquad & \xi^-_{ab} \; \longrightarrow \; \text{ghost graviton,}  \label{chi graviton limit}
\end{eqnarray}
we are left with the Einstein graviton mode and a ghost massive graviton ---unless we tune the couplings of the Lagrangian such that $m_g^2 \to \infty$ \cite{aspects}.

\subsubsection{Trace modes}
\label{subsec-Trace modes}

Since the trace equation \eqref{trace total ads} is already decoupled from the traceless (graviton) part, it is not difficult to undertake its mode decomposition. The only consideration is to take an analogue \textit{direct sum} prescription for the trace as we did in the traceless part: 
\begin{equation}
h := \phi_+ + \phi_- ~.
\label{trace decomposition}
\end{equation} 
For this we define
\begin{equation}
\phi_\pm := \beta_\pm (\bar\Box - m_{\phi_\mp}^2) h ~, \qquad \beta_\pm = - \frac{1}{m_{\phi_\pm}^2 - m_{\phi_\mp}^2} ~.
\label{eom scalar field phi}
\end{equation}
Therefore, the linearized equations of motion for the trace \eqref{trace total ads} become
\begin{equation} 
\pm \frac{1}{\kappa_\text{eff}^\text{trace}} (\bar\Box - m_{\phi_\pm}^2) \phi_\pm = \frac{T^L}{2} ~,
\label{scalar modes ads}
\end{equation}
where the effective Newton constant has an ugly expression,
\begin{eqnarray*}
\frac{1}{\kappa_\text{eff}^\text{trace}} & = & \frac{(d-1) \Lambda}{4} \left(\frac{(d-2) \left((d-2) \Lambda -m_g^2\right)}{\kappa _{\text{eff}} m_g^2 \left(d \Lambda +m_s^2\right)}+8 (d-1) d \Lambda  \mu +4 ((d-2) d+2) \Lambda  \nu \right) \times \\ [0.8em]
& & \!\!\!\!\!\!\!\!\!\!\!\!\! \sqrt{1-\frac{8 (d-2) \kappa _{\text{eff}} m_g^2 m_s^2 (4 (d-1) \mu +d \nu ) \left((2-d) \Lambda +m_g^2\right) \left(d \Lambda +m_s^2\right)}{\left(m_g^2 \left(4 \Lambda  \kappa _{\text{eff}} (2 (d-1) d \mu +((d-2) d+2) \nu ) \left(d \Lambda +m_s^2\right)-d+2\right)+(d-2)^2 \Lambda \right){}^2}} ~,
\end{eqnarray*}
depending on both $\mu$ and $\nu$. The alternate sign in \eqref{scalar modes ads} makes manifest the opposite behavior of $\phi_+$ and $\phi_-$; one of them is a ghost scalar field. If we take the limit $\mu, \nu \to 0$, then $m_{\phi_-} ^2 \rightarrow m_s ^2$, while $m_{\phi_+}^2 \rightarrow \infty$ and $\phi_+$ disappears from the spectrum (notice that $\kappa^* \to \infty$ in \eqref{kappas}). The effective Newton constant simplifies to
\begin{equation*}
\frac{1}{\kappa^\text{trace}_\text{eff}} \rightarrow \frac{(d-2) (d-1) \Lambda  ((d-2) \Lambda - m_g^2 )}{4 \kappa_\text{eff} m_g^2 (d \Lambda +m_s^2)} ~,
\end{equation*}
in full agreement with the results of $\mathcal{L}(Riemann)$ theories \cite{aspects}.

The detailed study of the propagating modes tell us that, contrary to the $\mathcal{L}(Riemann)$ case, there is no range of (non-vanishing) values of $\mu$ and/or $\nu$ leading to theories with a well-behaved spectrum. Ghosts are unavoidable and it is not possible to decouple them. There is no way to simultaneously have a positive effective Newton constant for all modes, $\kappa_{\xi_\pm}, \kappa_{\phi_\pm} > 0$ (this is particularly obvious for the latter, since $\kappa_{\phi_+} = - \kappa_{\phi_-}$), while avoiding tachyons in the spectrum, $m^2_{\xi_\pm}, m^2_{\phi_\pm} > 0$. The subset of theories with $\nu = 0$ admits a healthy graviton sector but we must further impose $\mu = 0$ if we want to avoid scalar ghosts. Therefore, $(\nabla Riemann)^2$ terms are unphysical for all possible couplings, contrary to what happens in $\mathcal{L}(Riemann)$ theories \cite{aspects}.

\subsection{A possible way-out involving higher-dimensional operators}

It is important to remark that our Lagrangian \eqref{nabla lagrangian final} is the most general one once we impose two requirements: i) having explicit derivatives of the curvature tensor, and ii) scaling as $[\mathcal{L}_{\nabla}] \sim E^6$. In other words, the claim that it is not possible to decouple the ghost particles unless we turn off both couplings, $\mu,\nu \to 0$, is only true once we demand both. We might want to examine other Lagrangians relaxing the latter condition; for instance:
\begin{equation}
\mathcal{L}_{\nabla} = \mu\, \nabla_a R \nabla^a R + \nu\, \nabla_a R_{bc} \nabla^a R^{bc} + \mu' \ell^2 R\, \nabla_a R \nabla^a R + \nu' \ell^2 R\, \nabla_a R_{bc} \nabla^a R^{bc} ~,
\label{lagrangian riem (nabla riem)2 example original}    
\end{equation}
where we introduced a length scale $\ell$ such that the couplings $\mu$, $\nu$, $\mu'$ and $\nu'$ have the same dimensions. Notice that the new terms have eight derivatives of the metric (equivalently, they scale as $E^8$), while the first two only have six derivatives. Therefore, condition ii) is not verified in \eqref{lagrangian riem (nabla riem)2 example original} and, in terms of a Wilsonian approach, this theory is mixing two different orders. However, they will contribute to the equations of motion at the same order. This is a tricky point in our discussion. There are terms which differ in their Wilsonian scaling while having the same number of derivatives in the equations of motion. This is actually well-known in the case of $\mathcal{L}(Riemann)$ theories, whose linearized equations of motion are fourth order regardless of the power of the contractions of the curvature tensors appearing in the Lagrangian.

Following the arguments discussed in Section \ref{sec-Linearization of the L(Riemann, nabla Riemann) contribution}, it is clear that the only contributions to the linearized equations of motion are those in which we vary twice with respect to $\nabla Riemann$, and evaluate everything else in the MSS background \eqref{riemann MSS}. Then, for the sake of studying the Lagrangian \eqref{lagrangian riem (nabla riem)2 example original} it is enough to consider
\begin{eqnarray}
\mathcal{L}_\nabla & \simeq & \left( \mu + \mu' \ell^2 \bar{R} \right) \nabla_a R \nabla^a R + \left( \nu + \nu' \ell^2 \bar{R} \right) \nabla_a R_{bc} \nabla^a R^{bc} \nonumber \\ [0.7em]
& = & \left( \mu + d(d-1) \mu' \ell^2 \Lambda \right) \nabla_a R \nabla^a R + \left( \nu + d(d-1) \nu' \ell^2 \Lambda \right) \nabla_a R_{bc} \nabla^a R^{bc} ~,
\end{eqnarray}
which is equivalent to \eqref{nabla lagrangian final} with the shifted couplings:
\begin{equation}
\mu \, \longrightarrow \mu + d(d-1) \mu' \ell^2 \Lambda ~, \qquad \nu \, \longrightarrow \nu + d(d-1) \nu' \ell^2 \Lambda ~.
\label{couplingshift}
\end{equation}
Therefore, the entire discussion on the spectrum performed in Sections \ref{subsec-Traceless modes} and \ref{subsec-Trace modes} carries on in the same way, just by replacing the coupling constants as in \eqref{couplingshift}. In the end, the requirement that the additional modes vanish reduces to imposing
\begin{equation}
\mu + d(d-1) \mu' \ell^2 \Lambda = \nu + d(d-1) \nu' \ell^2 \Lambda = 0 ~,
\label{vanishing-modes}
\end{equation}
which can be set in the Lagrangian \eqref{lagrangian riem (nabla riem)2 example original} without it becoming identically zero, thus making it possible to decouple the particles in a non-trivial way. This simple example serves to illustrate that there might exist indeed a higher-derivative gravity with sixth order Euler-Lagrange equations and an Einstein spectrum. However, it is an unnatural construct that is beyond the realm of a well-defined Effective Field Theory. It entangles the couplings of terms in the Lagrangian with different dimensional scaling, a relation that further demands the knowledge of the actual value of the cosmological constant in the MSS that serves as the vacuum. A honest computation of $\Lambda$ would require the introduction of all possible terms with the given scaling (in our previous example, $E^8$). In particular, we would need to consider altogether not only terms of the form $Riemann^4$ but also, say, $\nabla^2 Riemann\; \nabla^2 Riemann$, which we will scrutinize in the next Section.

In spite of the fact that this latter construction might deserve further attention ---since it could be useful as a toy-model for higher-curvature gravities involving explicit covariant derivatives of the Riemann tensor with the nice and non trivial property of having an Einsteinian spectrum---, we conclude that it is not possible to decouple the ghosts in a sixth order theory while preserving a well-defined EFT scheme. The very fact that the only way to achieve the decoupling involves mixing terms scaling differently in the Wilsonian expansion suggests an iterative procedure demanding going to all orders in the framework of an UV complete gravity theory.

All in all, two avenues are worth exploring: what happens if higher (covariant) derivatives of the Riemann tensor are introduced? And, in particular, what happens when an infinite series of these are brought into place?

\section{Spectrum of theories with higher derivatives of the Riemann tensor}
\label{sec-Spectrum of theories with higher derivatives of the Riemann tensor}

In this section we will discuss the propagating modes in theories of gravity whose Lagrangian is made of any combination of the Riemann tensor and its explicit covariant derivatives up to order $N$; \textit{i.e.}, $\mathcal{L} (Riemann, \nabla Riemann, \cdots , \nabla^N Riemann)$. To this end, we will study the linearized equations of motion. Let us consider perturbations around a MSS background $\bar{g}_{ab}$ with cosmological constant $\Lambda$, whose Riemann tensor is again $\bar{R}_{abcd} = 2 \Lambda\,g_{a[c|} g_{|d]b}$. Our aim is to describe the number of modes in the spectrum, which depends on the order of the derivatives in the Lagrangian. Rather than the Lagrangian itself, we will expand the integrand in the action, $L = \sqrt{|g|} \mathcal{L}$, to second order in the metric perturbations. It reads
\begin{eqnarray}
L & = & \bar{L} + \sqrt{|\bar{g}|} \, \bar{\mathcal{E}}^{ab} \delta g_{ab} + \Bigg[ \frac{\partial^2 \bar L}{\partial g_{ab} \partial g_{cd}} \delta g_{ab} \delta g_{cd} + \frac{\partial^2 \bar L}{\partial R_{abcd} \partial R_{efgh}} \delta R_{abcd} \delta R_{efgh} \nonumber \\ [0.9em]
& & \hspace{0.1cm} + \sum_{n = 1}^{N} \frac{\partial^2 \bar L}{\partial R_{abcd} \partial \nabla^n_{i_1 \cdots i_n} R_{efgh}} \delta R_{abcd} \delta \nabla^n_{i_1 \cdots i_n} R_{efgh} \label{eq General Lagrangian with arbitrary derivatives expanded} \\ [0.9em]
& & \hspace{0.1cm} + \sum_{m, n = 1}^{N} \frac{\partial^2 \bar L}{\partial \nabla^m_{i_1 \cdots i_m} R_{abcd} \partial \nabla^n_{j_1 \cdots j_n} R_{efgh}} \delta \nabla^m_{i_1 \cdots i_m} R_{abcd} \delta \nabla^n_{j_1 \cdots j_n} R_{efgh} \Bigg] + \cdots ~, \nonumber
\end{eqnarray}
where\footnote{Notice the symmetrization in the indices $i_1 \cdots i_n$, since the antisymmetric part could be rewritten in terms of covariant derivatives' commutators acting on curvature tensors, but these terms can be expressed simply as contractions of curvature tensors. Then, these produce contributions to the $\mathcal{L}(Riemann)$ part, so a redefinition of coupling constants can reabsorb them.} $\nabla^n_{i_1 \cdots i_n} := \nabla_{(i_1} \cdots \nabla_{i_n)}$  and the dots refer to higher functional derivatives of the Lagrangian. The bar over $L$ inside the brackets must be understood as variations followed by the evaluation on the MSS background. The first order term vanishes since the background is a solution of the equations of motion. The terms between brackets are of second order in the perturbation, $\delta g_{ab}$, thereby these are the ones that contribute to the linearized equations of motion when varying the action.

From this general expansion we see that the terms in the Lagrangian contributing to the linearized equations of motion are of the form
\begin{equation}
(Riemann)^l (\nabla^i Riemann)^m (\nabla^j Riemann)^n ~,
\label{eq Most general terms that contribute to the linearized EoMs}
\end{equation}
where $m i + n j$ must be an even number in order to be able to form scalars by contracting the indices. Also, since in \eqref{eq General Lagrangian with arbitrary derivatives expanded} we are taking first or second derivatives with respect to each $\nabla^i Riemann$ and then evaluating the result on the MSS background, only terms with $m + n \leq 2$ will effectively contribute to the spectrum.

Of course, $L$ can contain more terms than those in \eqref{eq Most general terms that contribute to the linearized EoMs}; \textit{e.g.}, we can consider corrections with $m + n > 2$ or contractions of the form $(\nabla^i Riemann) (\nabla^j Riemann) (\nabla^k Riemann)$ with $i \neq j \neq k$, which may certainly produce non-trivial contributions to the equations of motion.  However, they will vanish when linearized, since we must perform functional derivatives of the Lagrangian with respect to covariant derivatives of the Riemann tensor only twice, prior to evaluating it on the MSS background. Our claim therefore is that only terms that can be written as \eqref{eq Most general terms that contribute to the linearized EoMs} contribute to the linearized equations of motion. Also, noticing that
\begin{equation}
\delta R_{abcd} \sim \nabla^2 \delta g_{ab} ~, \qquad \delta \nabla_{i_1} \cdots \nabla_{i_k} R_{abcd} \sim \nabla^{k+2} \delta g_{ab} ~,
\end{equation}
it is straightforward to see from \eqref{eq General Lagrangian with arbitrary derivatives expanded} that a term like \eqref{eq Most general terms that contribute to the linearized EoMs} will produce contributions with \emph{at most} $m i + n j + 4$ derivatives of the metric perturbation (again, subject to the constraints that $mi + nj$ is even and $m + n \leq 2$).

The kind of corrections that we considered in the previous Section correspond to \eqref{eq Most general terms that contribute to the linearized EoMs} with $i = 1$, $j = 0$ and $n = 0$.  For these particular theories we argued that only the terms with $m = 0$ and $m = 2$ (in the notation of the current Section) contribute to the linearized equations of motion, whereas the different values of $l$ only modify the mass and the effective Newton constant of the various modes. In this Section, instead of working out the actual form of the linearized equations of motion, we will just focus on the simpler task of determining the order of the equations governing the perturbations (and, thus, on the number of modes) for more general theories of gravity.

\subsection{Linearized equations, field redefinition and ghosts}
\label{subsec-Linearized equations of motion in the general theory, field redefinition and ghosts}

Let us consider the most general theory with $2n+4$ derivatives,\footnote{Notice that all terms of the form \eqref{eq Most general terms that contribute to the linearized EoMs} with $m+n = 2$ can be rewritten, integrating by parts, as a combination of two different kinds of contributions: a set of $Riemann\,\nabla^{2s} Riemann$ terms, with different values of the exponent $s$, and ``residual'' terms of the form $Riemann^{r}$, which can be absorbed in the $\mathcal{L}(Riemann)$ part of the Lagrangian by redefining the couplings.}
\begin{eqnarray}
\label{general lagrangian order n}
\mathcal{L}(R_{ijkl}, \nabla^2 R_{ijkl}, \cdots ,  \nabla^{2n} R_{ijkl})=\sum_{s=0}^{n} \mathcal{L}_{\nabla^{2s}}=&  \\ [0.7em] \nonumber
=\mathcal{L}_{\nabla^0}(R_{ijkl}) +\mathcal{L}_{\nabla^2}(R_{ijkl} \nabla^2 R_{mnpq}) & + \cdots + \mathcal{L}_{\nabla^{2n}}(R_{ijkl} \nabla^{2n} R_{mnpq}) ~,
\end{eqnarray}
where $\nabla^{2n}$ amounts to the symmetric contraction of covariant derivatives and the previous expression must be understood, schematically, as an expansion in terms leading to fourth, sixth, \ldots, $(2n+4)^\text{th}$ order Euler-Lagrange equations. The (traceless) equations of motion can be written as
\begin{equation} 
\kappa_g^{-1} (\Box - 2 \Lambda) (\Box - 2 \Lambda - m_{g_1}^2) (\Box - 2 \Lambda - m_{g_2}^2) \cdots (\Box - 2 \Lambda - m_{g_{n+1}}^2) t_{ab} = \frac{1}{2} T^{L,\, \text{eff}}_{\langle a b \rangle} ~,
\label{traceless general eom}
\end{equation}
where $\kappa_g$ is some constant related with the coupling constants of the highest order terms in the Lagrangian, and $m_{g_i}^2$ are the masses of the $n+1$ massive gravitons that appear once we perform the appropriate fields' redefinition. Something similar happens to the trace part of the equations of motion:
\begin{equation} \label{trace general eom}
\kappa_{\phi}^{-1} (\Box - 2 \Lambda - m_{\phi_1}^2) (\Box - 2 \Lambda - m_{\phi_2}^2) \cdots (\Box - 2 \Lambda - m_{\phi_{n+1}}^2) h = \frac{1}{2} T^{L,\, \text{eff}} ~,
\end{equation}
where again $m_{\phi_i}^2$ would be the scalar field $\phi_i$'s mass and $\kappa_{\phi}$ is an overall constant related with the coupling constants of the highest order Lagrangian. At this point we can show that the canonical propagating degrees of freedom contained in \eqref{trace general eom} are such that there is always at least one ghost provided $m_{\phi_i}^2 \neq  m_{\phi_j}^2$ for all $i,j$. This generalizes the result of Section \ref{subsec-Trace modes}, which is nothing but the $n=1$ case.

To better see how the general procedure works, let us explore now the $n=2$ case, where we can use the Klein-Gordon operators $\hat{\Phi}_i := (\Box- m_{\phi_i}^2)$ to perform the fields redefinitions:
\begin{eqnarray}
\phi_1 &:=& \beta_1 \hat{\Phi}_2 \hat{\Phi}_3 h = \displaystyle\frac{1}{(m_{\phi_1}^2 - m_{\phi_2}^2) (m_{\phi_1}^2 - m_{\phi_3}^2)} \hat{\Phi}_2 \hat{\Phi}_3 h ~, \\ [0.7em]
\phi_2 &:=& \beta_2  \hat{\Phi}_1 \hat{\Phi}_3 h=\frac{1}{(m_{\phi_2}^2 - m_{\phi_1}^2) (m_{\phi_2}^2 - m_{\phi_3}^2)} \hat{\Phi}_1 \hat{\Phi}_3 h ~, \\ [0.7em]
\phi_3 &:=& \beta_3 \hat{\Phi}_1 \hat{\Phi}_2 h = \frac{1}{(m_{\phi_3}^2 - m_{\phi_1}^2) (m_{\phi_3}^2 - m_{\phi_2}^2)} \hat{\Phi}_1 \hat{\Phi}_2 h ~,
\end{eqnarray}
and, again, the coupling to matter of each scalar field would be given by
\begin{equation}
\kappa_{\phi_i} = \beta_i\,\kappa_{\phi} ~.
\label{scalarcouplings}
\end{equation}
It is straightforward to see that it is impossible to keep all $\beta_i$'s with the same sign, which means that there is always (at least) one ghost. Indeed, it is quite easy to write the result for the general case. The linearized equation of motion for the trace part can be written as
\begin{equation}
\kappa_{\phi}^{-1} \prod_{i=1}^{n+1} \hat{\Phi}_i\ h = \frac{1}{2}T^{L, \, \text{eff}} ~,
\end{equation}
and the fields' redefinition
\begin{equation}
\phi_i := \beta_i \prod_{k \neq i}^{n+1} \hat{\Phi}_k\ h ~, \qquad {\rm with} \qquad \beta_i^{-1} := \prod_{k \neq i}^{n+1} (m_{\phi_i}^2 - m_{\phi_k}^2) ~,
\end{equation}
show that each massive scalar fields' coupling is given by \eqref{scalarcouplings}. Given that $m_{\phi_i}^2 \neq m_{\phi_j}^2$ for all $i,j$, we can label the degrees of freedom such that $m_{\phi_{n+1}}^2 > \cdots > m_{\phi_{i}}^2 > \cdots > m_{\phi_{1}}^2$. Thereby, while $\beta_{n+1} > 0$, the mass gap $m_{\phi_{n+1}^2} > m_{\phi_{n}^2} $ flips the sign,
\begin{equation}
\beta_{n}^{-1} = (m_{\phi_{n}}^2 - m_{\phi_{n+1}}^2)(m_{\phi_{n}}^2 - m_{\phi_{n-1}}^2) \cdots (m_{\phi_{n}}^2 - m_{\phi_{1}}^2) < 0 ~,
\end{equation}
and in general $\text{sign}(\beta_k) = (-1)^{n+1-k}$. Taking this into account we conclude that in a general modified theory of gravity with $2n+4$ derivatives will have $n+1$ scalar fields, of which there will be either $\lfloor \frac{n+1}{2} \rfloor$ ghosts if $n+1$ is even with indepence of the rest of free parameters or $\lfloor \frac{n+1}{2} \rfloor  \pm  1$ ghosts if $n+1$ is odd, where the $\pm 1$ depends on the general couplings, $\Lambda$ and the spacetime dimension which determine the overall sign of $\kappa_{\phi}$ ($\lfloor x \rfloor$ is the integer part of $x$).

Regarding the graviton sector, some similar reasoning can be discussed, the general field redefinition to classify the different graviton modes of \eqref{traceless general eom} is given by: 

\begin{equation}
t^{(1)}_{ab} := \alpha_1  \prod_{k=1}^{n
+1} \hat{\mathcal{T}}_{k}  \, t_{ab} = \left( \frac{1}{  \prod_{k=2}^n m_{g_k}^2  } \right)  \prod_{j=1}^{n
+1} \hat{\mathcal{T}}_{j}  \, t_{ab} ~,
\end{equation}
where $\hat{\mathcal{T}}_{k}= \left( \bar{\Box}- 2 \Lambda - m_{g_k}^2 \right)$ and $t^{(1)}_{ab}$ is the Einstein graviton, which verifies the equation of motion
\begin{equation}
\frac{1}{\kappa_{g_1}} ( \bar{\Box}- 2 \Lambda) t^{(1)}_{ab} = T^{L, \, \text{eff}}_{\langle a b \rangle} ~.
\end{equation}
Focusing on the $n+1$ massive gravitons, the general field redefinition would be:
\begin{eqnarray}
t^{(j)}_{ab} &=& \alpha_j  ( \bar{\Box} - 2 \Lambda) \prod_{k \neq j}^{n+1} \hat{\mathcal{T}}_k\ t_{ab} \nonumber \\ [0.7em]
&=& \frac{1}{m_{g_j}^2} \prod_{i \neq j} \left(m_{g_i}^2 - m_{g_j}^2 \right)^{-1} (\bar{\Box}- 2 \Lambda) \prod_{k \neq j}^{n+1} \hat{\mathcal{T}}_k\ t_{ab} ~,
\end{eqnarray}
where $j = 2, \, \cdots, \, n+1$. Each of these modes verifies the equation
\begin{equation}
\frac{1}{\kappa_{g_j}} ( \bar{\Box}- 2 \Lambda - m_{g_j}^2) t^{(j)}_{ab} = T^{L,\, \text{eff}}_{\langle a b \rangle} ~,
\end{equation}
and again, if we assume no tachyonic particles ($m_{g_i}^2>0$) and no critical values (which we can express, on general grounds, as $m_{g_{n+1}}^2 > \cdots > m_{g_1}^2$), the same reasoning applies than in the scalar field sector due to the particular form of $\alpha_j$: in a theory with $2n+4$ derivatives we will have  $n+1$ massive gravitons of which there will be either $\frac{n+1}{2}$ ghosts if $n+1$ is even with independence of the rest of free parameters or $\frac{n}{2} \pm  1$ ghosts if $n+1$ is odd where the $\pm 1$ depends on the general couplings, $\Lambda$, and the spacetime dimensionality, which determine the overall sign of $\kappa_{g}$.

\section{Conclusions}

To sum up our results, in this letter we have discussed explicitly the equations of motion with sixth order derivatives (\textit{i.e.}, those corresponding to the $(\nabla Riemann)^2$ Lagrangian) on maximally symmetric spacetimes, showing the unavoidable presence of ghosts, in contrast with what has been found for theories built with general contractions of the curvature tensors, where it is indeed possible to decouple the pathological particles by suitable choices of the coupling constants \cite{ECG,GQTG,GQTGAllOrders}. This is a noteworthy result from the point of view of Effective Field Theory, since both kinds of corrections appear at the same order but they have substantially different behaviors in terms of their spectrum. So we conclude that the inclusion of explicit derivatives of curvature tensors in the Lagrangian, at least in a finite truncation on symmetric backgrounds, is ill-behaved.

This result is strongly supported by some other related articles \cite{Asorey, Modesto1, Biswas, Modesto2, Van, Belenchia} where it was shown, computing the propagator in flat spacetime, that ghosts are present in finite truncations of these higher derivative gravities. This work is consistent with those results, extending them to (A)dS spacetime as well (see, also, \cite{Biswas-2}).

We presented a counterexample to the previous claim, a higher-derivative gravity with sixth order Euler-Lagrange equations and an Einstein spectrum, but we argued that it lies beyond the realm of a well-defined Effective Field Theory. By entangling the coupling constants of terms scaling differently in the Lagrangian, it calls for the addition of all possible terms with that given scaling. This, in turn, prompted us to study the general case involving terms of the form $Riemann\,\nabla^{2n} Riemann$, which reintroduce the ghosts in the spectrum. The fact that to achieve the ghosts decoupling we are forced to mix terms in the Wilsonian expansion suggests, again, that consistence shall demand going to all orders in the higher curvature expansion involving explicit covariant derivatives of the Riemann tensor. This tantamount to the necessity of an UV complete gravity theory.

We have also computed the contribution to the boundary term when $(\nabla Riemann)^2$ terms appear in \eqref{varactionbdy}. A deeper study of the implications of this new term on the boundary can be of significance in some related holographic topics, like for instance the universal renormalization procedure in higher order gravities \cite{Araya} or the computation of asymptotic Noether charges.

More recently, the same conclusion has been obtained in the context of so-called dispersive CFT sum rules and bootstrap techniques \cite{Caron-Huot-1, Caron-Huot-2}. The punchline of all these (and other) works is that we cannot truncate the series and that all operators work together with the {\it right} coefficients; \textit{i.e.}, those admitting a healthy UV completion. This is in line with the findings of \cite{Biswas}, where a seemingly well-behaved (singularity free) higher derivative gravity involving covariant derivatives of the Riemann tensor is constructed, but it is non-local in nature, and cannot be written as a finite truncation of the series of higher derivative terms. It would be interesting to dig further into how stringent are these constraints in the construction of theories of gravity beyond Einstein's General Relativity.

\section*{Acknowledgements}

We would like to thank Alejandro Vilar for collaboration at early stages of this project and Sasha Zhiboedov for helpful comments. We would also like to thank Pablo Bueno, Pablo Cano and Robie Hennigar for their comments on the first version of this paper.
This work was supported by AEI-Spain (under project PID2020-114157GB-I00 and Unidad de Excelencia Mar\'\i a de Maeztu MDM-2016-0692), by Xunta de Galicia-Conseller\'\i a de Educaci\'on (Centro singular de investigaci\'on de Galicia accreditation 2019-2022, and project ED431C-2021/14), and by the European Union FEDER.
ARS is supported by the Spanish MECD fellowship FPU18/03719.
DVR is supported by Xunta de Galicia under the grant ED481A-2019/115.

\newpage
\appendix
\section{Decomposition of the traceless modes in $\mathcal{L}(Riemann)$ theories}
\label{sec-Appendix}

In decomposing our $t_{ab}$ tensor field we warned the reader that some subtleties must be taken into account. Let us first discuss them in the simpler case of $\mathcal{L}(Riemann)$ theories ({\it i.e.}, in the absence of $\nabla Riemann$ terms; $\mu = \nu = 0$), given that we already know the correct result \cite{aspects} and we can learn what is the problem with a wrong election. From \eqref{Epadmanabhan linealized traceless} and \eqref{Epadmanabhan trace} we can readily decouple the linearized traceless equation,
\begin{equation} 
\frac{1}{2 \kappa_{\text{eff}}\, m_g^2} (\bar{\Box} - 2 \Lambda) (\bar{\Box} - 2 \Lambda - m_g^2)\, t_{ab} = T_{\langle a b \rangle}^{L, \text{eff}} ~,
\label{traceless decouple L(Riemann) in (A)dS}
\end{equation}
which coincides with the result found in \cite{aspects}. We might want to do the field redefinition
\begin{equation}
\zeta_{ab} := \frac{\epsilon}{m_g^2} (\bar{\Box} - 2 \Lambda - m_g^2)\, t_{ab} ~,
\label{tab^m with no defined sign}
\end{equation}
where we introduced a floating (yet undefined) sign, $\epsilon = \pm 1$. It trivially fulfills the equation of motion
\begin{equation}
(\bar{\Box} - 2 \Lambda)\, \zeta_{ab} = 2 \epsilon \kappa_{\text{eff}}\, T_{\langle a b \rangle}^{L, \text{eff}} ~,
\label{wrong-sign}
\end{equation}
given that $\epsilon^2 = 1$. We already see that the correct sign for the Einstein graviton coupling occurs only if we take $\epsilon = -1$. Let us ignore this for the moment and focus on the massive graviton mode, which could be defined via
\begin{equation}
\xi_{ab} := t_{ab} + \lambda\, \zeta_{ab} = \frac{\epsilon}{m_g^2} \Big[ (\epsilon - \lambda) m_g^2 + \lambda (\bar{\Box} - 2 \Lambda) \Big]\, t_{ab} ~,
\label{mass-remnant}
\end{equation}
where $\lambda$ is a sign that should be fixed as $\lambda = \epsilon$ in order to kill the mass remnant in \eqref{mass-remnant}. Thereby,
\begin{equation}
\xi_{ab} = \frac{1}{m_g^{2}} (\bar{\Box} - 2 \Lambda)\, t_{ab} ~,
\end{equation}
so that its equation of motion reads:
\begin{equation}
(\bar{\Box} - 2 \Lambda - m_g^2)\, \xi_{ab} = 2 \kappa_{\text{eff}}\, T_{\langle a b \rangle}^{L, \text{eff}} ~.
\end{equation}
In summary, the sign of the coupling to the stress-energy tensor tells us that we are left with a ghost massive graviton and an apparent ambiguity in the Einstein mode. Had we chosen $\epsilon = +1$, the \textit{wrong} sign, the mode decomposition of $t_{ab}$ \eqref{mass-remnant} would have been $t_{ab} = \xi_{ab} - \zeta_{ab}$. Thereby, we might interpret that the wrong sign in the coupling of the massless graviton to the stress-energy tensor in \eqref{wrong-sign} merely comes from the minus sign of the tensor decomposition. However this is just an apparent issue based on the fact that we are reading the causal behavior directly from our equations of motion without taking into account the action normalization encoded in that particular sign choice.

If we, on the other hand, expand the action in modes allowing for relative signs between them, what we are implicitly assuming is the fact that the prescription to couple matter fields with gravity is mode dependent. This is manifestly unphysical. Thereby, we must impose that the whole decomposition is a direct sum of all modes, $t_{ab} = t_{ab}^{(m)} + t_{ab}^{(M)}$, which happens for $\epsilon = -1$.

\end{document}